\documentstyle[aps,prl,epsf,preprint]{revtex}
\begin{document}
\widetext
\title{Phase ordering kinetics of the Bose gas}
\author{Kedar Damle, Satya N. Majumdar and Subir Sachdev}
\address{Department of Physics, P.O. Box 208120, Yale University,
New Haven, CT 06520-8120}
\date{\today}
\maketitle
\begin{abstract}
We study the approach to equilibrium of a Bose gas to a superfluid state. We
point out that dynamic scaling, characteristic of far from equilibrium
phase-ordering systems, should hold. We stress the importance of a
non-dissipative Josephson precession term in driving
the system to a new universality class. A model of coarsening in
dimension $d=2$, involving a quench between two temperatures below the
equilibrium  superfluid transition temperature ($T_c$), is exactly solved and
demonstrates the relevance of the Josephson term. Numerical
results on quenches from above $T_c$ in $d=2,3$ provide evidence for
the scaling picture postulated.

\end{abstract}
\pacs{PACS numbers:75.10.Hk, 64.60.Cn, 64.60.My, 64.70.Md}

\widetext

The recent observation of Bose condensation in neutral, trapped atomic
gases~\cite{wieman} and excitons in $\rm{Cu}_2 \rm{0}$~\cite{wolfe} opens up
exciting possibilities on experimental studies of time-dependent non-equilibrium
phenomena
in a heretofore inaccessible regime. In particular, an issue which could be
experimentally investigated, and which we shall address theoretically in this
paper is the following---upon quenching a Bose gas to a final temperature
($T$) below $T_c$, how does the condensate density
grow with time before attaining its final equilibrium value? A few recent
papers~\cite{otherbose,kagan} have addressed
just this question, but they have focussed on the early time (on the order
of a few
collision times), non-universal dynamics.
However, as has also been noted recently in Ref~\cite{KS},
the interesting
experimental questions are instead associated with the long-time dynamics
involving
the coarsening of the Bose condensate order parameter. This dynamics is
``universal'' in a sense that will be clarified below.

A natural and precise language for describing the evolution of the condensate is
offered by recent developments in the theory of phase-ordering dynamics in
dissipative classical spin systems, as reviewed in the article by
Bray~\cite{bray}. In this theory, one considers the evolution of a classical
spin system after a rapid quench from some high $T$ to a low
$T$ in the ordered phase. The dynamics is assumed to be purely relaxational, and
each spin simply moves along the steepest downhill direction in its
instantaneous
energy landscape. Locally ordered regions will appear immediately after the
quench, but the orientation of the spins in each region will be different.
The coarsening process is then one of alignment of neighboring regions, usually
controlled by the motion and annihilation of defects (domain walls for Ising
spins, vortices for $XY$ spins etc.). A key step in the theory is the
introduction of a single length scale, $l (t)$, a monotonically increasing
function of the time
$t$, which is about the size of a typical ordered domain at time $t$.
Provided $l
(t)$ is greater than microscopic length scales, like the range of
interactions or
the lattice spacing, it is believed
that the late stage morphology of the system is completely characterized  by
$l (t)$, and is independent of microscopic details,
{\em i.e.} it is universal. This morphology is characterized by various time
dependent correlation functions which exhibit universal scaling behavior.

We turn then to the Bose gas. The order parameter in this case is the boson
annihilation field $\psi(r, t)$ (where $r$ is a spatial co-ordinate); the phase
of the expectation value of $\psi$ is aligned across the system in the
equilibrium Bose-condensed state. A key point is that after relatively
few atomic collisions, once the domain size $l(t)$ is large enough ({\em e.g.\/}
larger than the de Broglie wavelength), it is permissible~\cite{kagan} to
treat
$\psi (r, t)$ as a {\em classical\/} field which obeys Hamilton-Jacobi
equations of
motion (for a related discussion on the emergence of classical dynamics in the
equilibrium properties of an antiferromagnet, see Ref~\cite{CHN}). It must
be kept
in mind that it is only the equations of motion for the collective order
parameter
which are classical---the very existence of the complex order parameter is due
entirely to quantum mechanics, and the fact that there is a wavefunction for the
condensate.

An important property of the equations of motion for $\psi$, discussed below, is
that they are not simply relaxational. Instead, they contain
non-dissipative, kinematical ``streaming'' or ``Poisson bracket''
terms~\cite{halphoh}. One such term is responsible for the Josephson precession
of the phase of $\psi$ at a rate determined by the local chemical potential.
A central objective of this paper is to understand the consequences of such
terms on the phase-ordering theories of Ref~\cite{bray}. We will argue that
the Josephson term constitutes a relevant perturbation on the dynamics and
that the
resulting coarsening process belongs to a new universality class.
Specifically, in the remainder of the paper we will ({\em i})
exactly solve a model of a $d=2$ Bose gas always in contact with a
reservoir, where the temperature of the reservoir is suddenly switched between
two temperature below $T_c$~\cite{RB}; ({\em ii}) present numerical
results on the time evolution
of an isolated Bose gas in $d=2,3$, where the initial state has no superfluid
fraction, while the final state is superfluid.

We will begin by considering a solvable coarsening problem which illustrates the
possible consequences of the Josephson term in a simple setting.
A $d=2$ Bose gas is superfluid for $T < T_{KT}$, the well known
Kosterlitz-Thouless phase transition temperature; consider the phase ordering
process in which the Bose gas is moved at time $t=0$ from contact with a
reservoir at an initial
$T = T_i$, to a reservoir with a final
$T=T_f$, such that
$T_f < T_i < T_{KT}$; a similar quench was considered in Ref~\cite{RB} for the
purely dissipative $XY$ model. In the long-time limit, all vortices and
fluctuations in the amplitude of
$\psi$ can be neglected, and we may
parametrize $\psi = e^{i \phi}$.
The free energy density in the purely dissipative $XY$ model~\cite{RB} is now
determined simply by the gradients of the phase $\sim (\nabla \phi)^2$. In the
case of the Bose gas, it is also necessary to take the conserved number density
into account. Let $m$ be proportional to the deviation of the particle density
from its mean value; then the free energy density we shall work with is
\begin{equation}
{\cal F} =\frac{1}{2}\int d^2 r [(\nabla
\phi)^2+m^2].
\end{equation}
We have rescaled spatial co-ordinates and $m$ to obtain convenient coefficients
in ${\cal F}$. The Josephson precession term, whose effects we wish to
study, is
contained in the Poisson bracket
\begin{equation}
\{ m(r), \phi(r^{\prime}) \} = g_0 \delta(r - r^{\prime}),
\end{equation}
where $g_0$ is a constant.
The method reviewed in Ref~\cite{halphoh} now leads to the
{\em linear} equations of motion~\cite{nelson}
\begin{eqnarray}
{{\partial \phi}\over {\partial t}}&=&\Gamma_0 {\nabla}^2\phi+g_0 m+\theta,
\nonumber \\
{{\partial m}\over {\partial t}}&=&\lambda_0 {\nabla}^2m+g_0
{\nabla}^2\phi +\zeta
\label{mphieqn}
\end {eqnarray}
where the coefficients $\Gamma_0, \lambda_0 > 0
$ represent the dissipation arising from the coupling of the system to the
reservoir. The effects of the reservoir are also contained in the Gaussian
noise
sources $\theta$ and $\zeta$ with zero mean and (for $t>0$) correlations
appropriate to $T=T_f$:
$\langle \theta (r,t) \theta (r' ,t')\rangle=2\Gamma_{0}T_f\delta
(r-r')\delta (t-t')$,
$\langle \zeta (r,t) \zeta (r' ,t' )\rangle=-2\lambda_{0}T_f
{\nabla}^2\delta (r-r')\delta (t-t')$ and
$\langle \zeta (r,t) \theta (r' ,t' )\rangle=0$
($k_B = 1$).
 The equations
(\ref{mphieqn}) are linear, can be easily integrated, and all correlations
can be
computed exactly.

Let us first recall the structure of the solutions expected from naive
scaling~\cite{bray} for
$d \geq~2$. For our models we expect the domain size $l (t) \sim t^{1/z}$
where $z$
is a non-equilibrium dynamic exponent.
We consider the behavior of two correlation
functions: ({\em i}) The equal-time correlator $G (r, t) = \langle \psi^{\ast}
(r,t) \psi (0, t)\rangle $ (the $k=0$ component of the spatial Fourier
transform of
$G$ is proportional to the condensate fraction). Under scaling we expect
for large
$r$ and $t$, 
$G(r, t) \sim r^{-\eta_f} g (r/t^{1/z})$ where $g$ is a universal scaling
function, $\eta_f = 0$ for
$d >2$, while in $d=2$, $\eta_f$ ($=T_f /2 \pi$ for ${\cal F}_1$) is the
equilibrium exponent of the quasi-long-range order at $T=T_f$.
({\em ii}) The unequal time correlation function $C(r,t) = \langle
\psi^{\ast} (r,t ) \psi (0,0) \rangle$ for which we expect for large $r$ and
$t$, $C(r,t)
\sim t^{-\lambda/z} f(r/ t^{1/z})$ where $f$ is a universal scaling
function, and
$\lambda$ is a new dynamic exponent.

It turns out that our model ${\cal F}$ does not
completely obey the simple scaling hypotheses as stated above. This becomes
clear upon
considering the two-time correlation, $C$,
which turns out to depend upon {\em two} large length scales, $\l_1 (t) \sim (a
t)^{1/2}$ and
$l_2 (t) \sim g_0 t$ (with $a = (\Gamma_0 + \lambda)/2$): it obeys the scaling
form $C(r,t) \sim t^{-(3
\eta_i + \eta_f )/4} \tilde{f} ( r/ (a t)^{1/2} , r/(g_0 t))$ (where
$\eta_i = T_i
/2 \pi$). The dependence of these scales on
$g_0$ suggests that $g_0$ is a relevant perturbation with renormalization group
eigenvalue 1, in the language of Ref~\cite{bray}. The scaling function
$\tilde{f}$ was determined to be
\begin{equation}
\tilde{f} (x_1 ,x_2 )=\exp [-{\eta}_i\int_0^{\infty}{dy\over y}\{1-J_0(y)\}
\cos ({y/x_2 })
 e^{-y^2/x_1^2}].
\end{equation}
For $r \sim l_1 (t)$, we use $\tilde{f} ( x_1 , x_2 \rightarrow 0 ) =1$:
this shows that the autocorrelation $C(0,t)\sim t^{-
(3\eta_i+\eta_f)/4}$ in contrast to the result in the model of
Ref~\cite{RB}
$C(0,t)\sim t^{-(\eta_i+\eta_f )/4}$.
On the contrary, one could insist on a scaling picture using only the single
larger length
scale $r\sim l_2 (t)$, and would then need $\tilde{f} (x_1 \rightarrow \infty,
x_2 )$
which equals $[1+\sqrt{1-x_2^2}]^{-{\eta}_i}$ for $x_2<1$ and equals
$x_2^{-\eta_i}$ for $x_2\geq1$.
It can also be checked that
one recovers the initial equal time equilibrium result for $C(r,t)$ when
$r\rightarrow \infty
$ with $t$ large but fixed. We also note that
the relevance of $g_0$ is evident in the autocorrelations of $m$.
We find $\langle m(0,t) m(0, 0) \rangle \sim
(1/t) f_1 (g_0 {\sqrt {t / a}})$ where
\begin{equation}
f_1(\tau )=4{\pi}^2{\eta}_i[1-\int_0^{\infty}\sin y e^{-y^2/{2\tau^2}}dy];
\end{equation}
clearly, for $g_0 =0$, this autocorrelator decays as $1/t$ for large $t$,
while for
nonzero $g_0$ it decays faster as $t^{-2}$. Finally, results on the
equal-time  $\psi$ correlator, $G$. It has a crossover at a time $t_1
\sim \tilde{a} /g_0^2$ with $\tilde{a} = |\Gamma_0 - \lambda_0|$; this time is
similar to the crossover time in  $\langle m(0,t) m(0, 0) \rangle$, except that
$\tilde{a}$ has replaced $a$.  Both for $t \ll t_1$ and for $t \gg t_1$,
$G$ obeys
a scaling form similar to that obtained in the relaxational model of
Ref~\cite{RB}
(which has $g_0 = 0$):
$G( r, t) \sim  r^{-\eta_f} g(r/(\gamma t)^{1/2})$ where
$g$ is a scaling function described in Ref~\cite{RB}; however, the rate
$\gamma =
\Gamma_0$ for $t \ll t_1$ and $\gamma = a$ for $t \gg t_1$.

While this phase only model-$\cal F$ is not relevant for studying quenches
from above the transition temperature (since it neglects the non-linear
terms), the exact solution of this linear model is quite instructive.
It clearly emphasizes the
importance of the nondissipative Josephson coupling term. In fact as seen
above, the presence of this term ($g_0\neq 0$) changes the universality
class of the system. Thus it is reasonable to expect that even for quenches
from above the transition temperature, this term would play an important
role. In this case, after the quench the system has defects (e.g., vortices
in $2$-d). As time progresses, these defects move around and annihilate
each other and the system becomes more and more ordered. To study this
coarsening process that proceeds via the annealing of defects it
is necessary to study the evolution of both the phase and amplitude of
$\psi$. This growth of long range order in the system can be studied in
two different ways. In one case one considers the
time evolution of an {\em isolated\/} Bose gas, not in contact with a
reservoir. Though the dynamics in this case is nondissipative, the system still
exhibits irreversible approach to equilibrium. In the other case, the Bose gas
is in contact with a heat bath.
These are the analogues of micro-canonical and canonical ensembles
in equilibrium statistical mechanics. It is reasonable to expect that both
descriptions would lead to the same results for universal scaling
properties. Most previous studies on coarsening have been done in the
``canonical" ensemble. However in this paper, we use the ``microcanonical"
approach. To the best of our knowledge, this approach has never been
used before to study coarsening in any system. As we will see below, the
dynamics in the ``microcanonical" ensemble is completely specified by
the Hamiltonian of the system with no additional phenomenological parameters.
The ``canonical" dynamics, on the other hand, needs several phenomenological
constants as input parameters.

For the isolated Bose gas (``microcanonical" ensemble), an excellent
approximation for the total energy of an order
parameter configuration $\psi (r,t)$ is
$
{\cal H}  =\int d^d r \left[|\nabla
\psi|^2  + \frac{u}{2} |\psi|^4 \right],
$
where we have rescaled lengths to make
the coefficient of the gradient term unity, and $u>0$ is the two-particle
$T$-matrix at
low momentum. The Hamilton-Jacobi equation of motion for $\psi$ follows from the
Poisson bracket $\{ \psi , \psi^{\ast} \} = i$
\begin{equation}
i{{\partial \psi}\over {\partial t}}=[-{\nabla}^2+u|\psi|^2]\psi ,
\label{1}
\end{equation}
and is the well-known~\cite{gp} Gross-Pitaevski (GP) or non-linear Schroedinger
equation. We can also add a quadratic $|\psi|^2$ term to ${\cal H}$, and it
leads
to term linear in $\psi$ in the GP equation; however this linear term can be
eliminated by an innocuous global phase change in $\psi$. The GP equation
conserves the total number of particles $\int d^d r |\psi|^2$, the total
momentum, and
${\cal H}$, and hence there is no global dissipation of energy.
Nevertheless, in the
thermodynamic limit, the GP equation does display irreversible coarsening,
as will
be abundantly clear from our numerical results to be described later: a random
initial state with a negligible number of particles in the zero momentum ($k$)
state (i.e., short range initial correlations), evolves eventually to a state
with a condensate fraction equal to that
expected at equilibrium in the microcanonical ensemble at the total energy
of the initial state.

In the ``canonical" approach on the other hand, it is permissible to add
dissipative terms to the equation of motion of $\psi$. A simple additional
damping term to the GP equation leads to a model expected to be in the same
universality
class of the so-called Model~A~\cite{halphoh,bray}; this model is however not
acceptable: it violates local conservation
of the particle density, and, as discussed near (\ref{mphieqn}), it is
necessary~\cite{halphoh,siggia} to introduce the density fluctuation field,
$m(r,t)$; the value of
$|\psi (r,t)|^2$ is then the contribution to the particle density from low
momentum
states, while
$m(r,t)$ represents the density fluctuation from all states; the Poisson bracket
in this case is
$\{ m(r) , \psi(r^{\prime}) \} = i g_0 \psi(r) \delta(r-r^{\prime})$.
This is model F in the language of Ref~\cite{halphoh} (see also \cite{momcons}). 
Note that the strength of the crucial precession term in the dynamics is
controlled by $g_0$ which is an adjustable phenomenological parameter (however,
in the Hamiltonian dynamics considered earlier, there is no such freedom).
Numerical study of coarsening using model F could thus be complicated by
crossover effects associated with the adjustable value of $g_0$ ($g_0=0$
corresponds to the purely dissipative model-A dynamics, which is clearly in a
different universality class). 

We therefore restrict our numerical study here to the GP model. 
All of the numerical results obtained so far are
consistent with the simplest naive scaling hypotheses described earlier, and do
not require the introduction of two length scales, as was
necessary in the linear model above (though we have not yet obtained numerical
results on unequal time
correlations, for which the linear model ${\cal F}$ clearly displayed two
length scales).  We will present results both in $d=2$ and $d=3$. The
$d=2$ system allowed us to study
larger sizes with better finite-size scaling properties.

We discretized (\ref{1}) on a lattice, and
integrated in time using a FFT-based algorithm which conserved energy and
particle number to a high accuracy. We work in units where the lattice
spacing is unity and choose the scale of the lattice field to make
the density one. We set $u$ to be approximately $0.25$ so that we are
considering a dilute gas.
We choose an ensemble of initial conditions with a narrow distribution of
energy,
whose width goes to zero in the thermodynamic limit. We assign initial
values to the
Fourier components
$\psi (k,0)$ as follows: $\psi (k,0)=\sqrt {n_0(k)}\exp[i\phi (k)]$ where
the $\phi (k)$'s are independent random variables chosen from a uniform
distribution with range $[0,2\pi]$ and the function $n_0(k)$ is chosen to
ensure that initial real-space correlations are short-ranged (corresponding to
a ``high temperature" configuration) while still having low enough energy so
that the equilibrium state corresponding to this energy is superfluid. Though
the ensemble of initial conditions is  not strictly the
Gibbs distribution  at any temperature, it is however expected that the precise
details of the initial conditions do not matter for the late time
universal properties as long as the initial correlations are short ranged.
 
More specifically we chose $n_0(k)= c/((\epsilon (k) +{\mu}_1)
(\exp[(\epsilon (k)-{\mu}_2)/ T]+1))$ where $\epsilon (k)$ is the 
fourier representation of the lattice version of the Laplacian and $c$
sets the overall scale of $n_0(k)$.
Here one chooses the parameters ${\mu}_1$, ${\mu}_2$ and $T$ to 
achieve the appropriate trade-off between energy and correlation length.
Note that this careful choice of initial conditions is needed as the GP
equation does not have any explicit dissipation and the system evolves in phase
space on a constant energy surface.

We used finite-size scaling to model the results in a finite system of linear
dimension $L$: it predicts a scaling form
$G(r,t)=L^{-\eta} \Phi_G [r/L,t/L^z]$ for the equal time
correlation function. In $d=3$ the exponent $\eta =0$, while in $d=2$, it
is associated with the final equilibrium state and varies continuously with
temperature. The structure factor, $S(k,t)$, is obtained by a spatial Fourier
transform of
$G(r,t)$, and the number of particles in the $k=0$ mode is clearly $S(0,t)$;
the latter should satisfy
$S(0,t)
\sim L^{2-\eta} \Phi [t/L^z]$ in $d=2$ and $S(0,t)
\sim L^{3} \Phi [t/L^z]$ in $d=3$.
The scaling
function
$\Phi$ goes to a constant for $t \gg L^{z}$ and the system attains equilibrium
after a time
$t\sim L^z$.

\begin{figure}
\epsfxsize=5.8in
\centerline{\epsffile{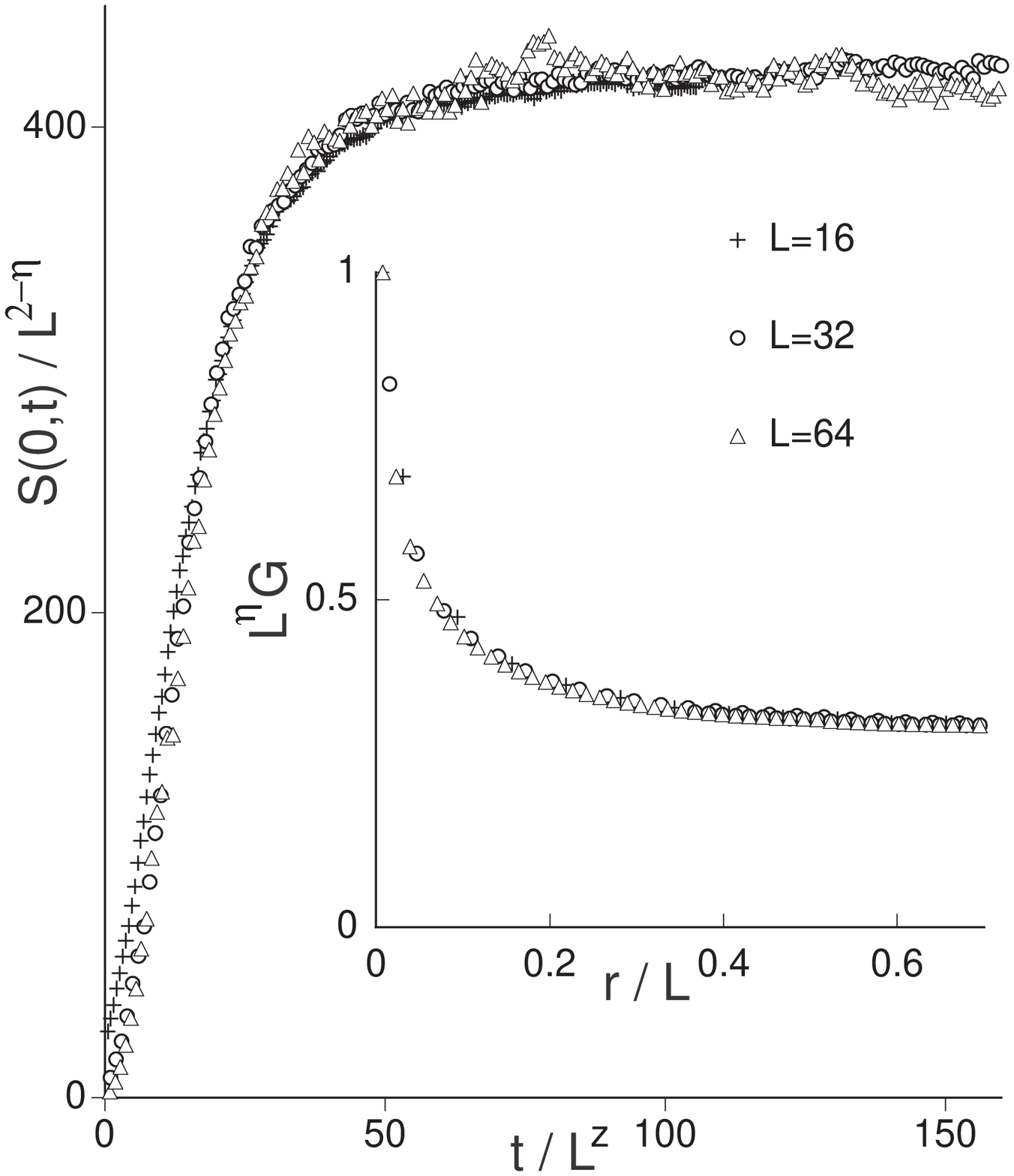}}
\vspace{0.5in}
\caption{Numerical results from the simulation of the GP equation in $d=2$.
The number of particles in the zero momentum state is $S(0,t)$ and the figure
shows its scaling properties as a function of system size, $L$, and time, $t$.
The inset shows the scaling of the equilibrium equal time correlation function,
$G(r, t \rightarrow \infty)$.
The best scaling collapse was obtained in both plots for $\eta \approx
0.27$ and $z
\approx 1.1$. The scale of all axes (except the values of $r/L$) are arbitrary.}
\label{f1}
\end{figure}

\begin{figure}
\epsfxsize=5.8in
\centerline{\epsffile{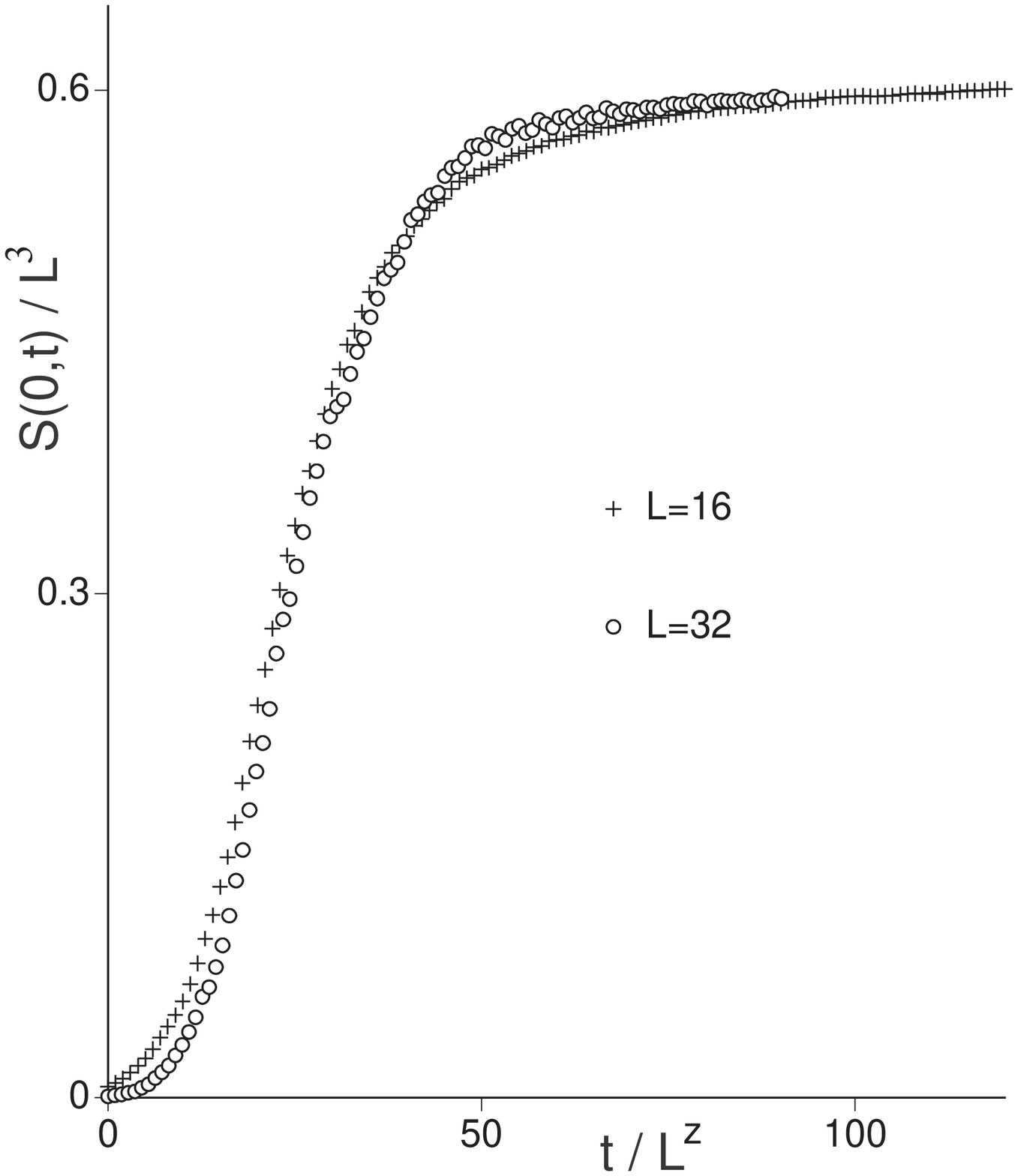}}
\vspace{0.5in}
\caption{Numerical results for the GP equation in $d=3$. The notation is as in
Fig.~\protect\ref{f1}, with the exponent $z \approx 1.15$.}
\label{f2}
\end{figure}

Results for $d=2$ are shown in Fig~\ref{f1}. We performed
finite size scaling analysis for
$L=16$, $32$
and $64$ and found reasonable data collapse with $\eta\approx 0.27$ and
$z\approx
1.1$. The value of $\eta$ indicates that we are at a non-zero
temperature close to  $T_{KT}$; strictly speaking we must have $\eta < 1/4$,
but the value of $\eta$ is relatively $T$ independent near $T_{KT}$, and the
discrepancy is within our numerical errors. The value of $z$ is
in sharp contrast to the $z=2$ (with logarithmic corrections) result obtained
by various groups~\cite{YU,MO} for the purely dissipative Model~A
dynamics~\cite{halphoh} (obtained from
Model~F by setting $g_0 = 0$ and ignoring $m$) of classical
$XY$ spins. Although we have determined the value of $z$ for a quench to a
particular temperature $T_f$, we expect that $z$ is
the same for all $0<T_f<T_{KT}$.
Results for $d=3$ are shown in Fig~\ref{f2} for linear sizes $L=16,32$. The data
collapse is not as good as that in $d=2$, but again we obtained a $z
\approx 1.1$. 

Thus our numerical results, both in $d=2$ and $3$, are consistent
with a value of $z=1$, which is also the result suggested by the exact
calculation in the phase only model. 

Finally, we close with some physical discussion 
on reasons for the difference between the GP model, and quenches in the purely
dissipative
Model~A~\cite{YU,MO}. The dynamics in the GP model proceeds
via the annihilation of nearby vortex-antivortex pairs (in $d=2$) as in Model~A.
However there is an
important difference between the two in the details of the vortex motion. In
Model~A, oppositely charged vortices attract each other with a force that
falls off as the inverse of their separation (apart from logarithmic
corrections). Since the
dynamics is overdamped, this implies $l(t)\sim t^{1/2}$. In the GP model, on the
other hand, the situation is more complex. In addition to vortices, the system
also has a propagating ``spin-wave" mode arising from the streaming terms.
A pair of oppositely charged vortices, apart from attracting each other, also
interacts with the spin-wave background. In addition, it experiences a Magnus
force which causes the pair to move with uniform velocity in a
direction perpendicular to the line joining them. These qualitative
differences in the nature of the defect dynamics change the universality
class of the coarsening process.

In summary, we have presented evidence, both analytical and numerical, that
the phase-ordering dynamics of the Bose gas belongs to a new universality
class. A particular conclusion of this work is that the condensate density
of the Bose gas, following a sudden quench from the normal to the superfluid
phase in dimensions $d\geq 2$, will grow at late times as $t^{d/z}$. We have
presented evidence, both analytical and numerical, that $z=1$.

We thank D. Kleppner for stimulating our interest in this problem, and
M.P.A. Fisher and A. Bhattacharya for useful discussions.
This research was supported by NSF Grants DMR-92-24290 and DMR-91-20525.



\begin{references}
\bibitem{momcons} It may also be necessary to introduce additional fields 
to account for other conserved quantities: a momentum density as in model-H
or an energy density as in model-C of Ref. $8$.
\end{references}
\end{document}